\documentclass[12pt]{article}
\usepackage[english]{babel}
\usepackage{amsmath,amsthm,amssymb,amsfonts,latexsym,url}
\usepackage{setspace,color}
\newtheorem{theorem}{Theorem}

\begin{document}

\title{Continua  with non-local constitutive laws: \\
exploitation of entropy inequality}

\author{M.~Gorgone, F.~Oliveri and P.~Rogolino\\
\ \\
{\footnotesize Department of Mathematical and Computer Sciences,}\\
{\footnotesize Physical Sciences and Earth Sciences, University of Messina}\\
{\footnotesize Viale F. Stagno d'Alcontres 31, 98166 Messina, Italy}\\
{\footnotesize mgorgone@unime.it; foliveri@unime.it; progolino@unime.it}
}

\date{Published in \textit{Int. J. Non-Linear Mech.} \textbf{126}, 103573 (2020).}

\maketitle

\begin{abstract}
In this paper, we consider a system of balance laws sufficiently general to contain the  equations  describing the thermomechanics of a one-dimensional continuum; this system involves some constitutive functions depending on the
elements of the so called state space assumed  to contain the spatial gradients of some of the unknown fields. The compatibility of the constitutive equations with an entropy-like principle is considered via an  extended Liu procedure by using as constraints both the balance equations and some of their gradient extensions.
This procedure is then applied to the equations of a fluid whose description involves an internal variable and 
first order non-local constitutive relations, and to a Korteweg fluid with second order non-localities.
In both cases, the restrictions placed by an entropy inequality are solved, and an explicit solution for the constitutive equations is provided.
\end{abstract}


\section{Introduction}

In continuum thermodynamics, the constitutive theories are based, besides some general invariance principles, on the second law of thermodynamics, which states that in every admissible process the entropy production has to be non-negative \cite{Truesdell}.

A rigorous procedure for the exploitation of the entropy principle has been developed for the first time in 1963 by Coleman and Noll \cite{Coleman-Noll}, and later by Coleman and Mizel \cite{Coleman-Mizel}. In both papers, the authors assumed the unbalance of entropy 
in the classical form, say the Clausius-Duhem inequality; in this inequality, the entropy flux  is taken as the ratio between the heat flux and the absolute temperature. Later on, M\"uller \cite{Muller} proposed an extension of the entropy inequality, allowing a more general expression for the entropy flux, thus obtaining the thermodynamic compatibility for wider classes of materials. A slightly different approach has been applied by other authors who accepted the classical Clausius-Duhem inequality, but proposed a more general form of the local balance of energy \cite{Gurtin,Dunn-Serrin,Dunn}. Furthermore,  in 1972, Liu  \cite{Liu} developed a different procedure for the analysis of the entropy principle, based on the method of Lagrange multipliers.

In all these papers, the basic assumption is that the second law of thermodynamics restricts the constitutive equations and not the thermodynamic processes. Hence, the constitutive relations are required to be such that the entropy inequality be satisfied for all solutions of the thermodynamic field equations. This assumption is purely mathematical and from a physical point of view may have two
different interpretations \cite{Muschik-Ehrentraut}:
\begin{itemize}
\item all solutions of the balance equations have to satisfy the second law;
\item there are solutions of the balance equations which satisfy the second law, and other ones which do not. 
\end{itemize}

The first interpretation requires that the constitutive equations must be assigned in such a way the entropy inequality is satisfied along arbitrary processes, whereas the second one means that we have to exclude from the set of solutions of the balance equations those which are not physically achievable, since they do not satisfy the second law of thermodynamics. In \cite{Muschik-Ehrentraut},  the authors
proposed a way to choose between the two statements through an \emph{amendment to the second law}, by expliciting the nearly self-evident, but never precisely formulated, postulate that there are no reversible process directed towards non-equilibrium. 
By means of this amendment, they were able to prove that, necessarily, the second law of thermodynamics restricts the constitutive equations and not the processes. Such a result justifies, from the physical point of view, the approach to the exploitation of second law 
through  Coleman-Noll and Liu procedures (see also \cite{Muschik,Muschik-Papenfuss-Ehrentraut}).

In a series of papers \cite{Cimmelli-2007,CST-JMP-2009,CST-PRSA-2010,COT-JMP-2011},
the two classical Coleman-Noll and Liu procedures have been extended in order to use as constraints in
the entropy inequality both the balance equations and their gradient extensions up to a suitable finite order. 
This approach, successfully used in many applications of physical interest (see, for instance, 
\cite{CST-JNET-2010,COP-Elasticity-2011,COP-IJNLM-2013,COP-CMT-2015,OPR-2016,CGOP-miscele-2020}), 
revealed essential in order to ensure the compatibility of non-local constitutive relations with second law of thermodynamics 
without modifying \emph{a priori} the entropy inequality or the energy balance through the introduction of extra-terms.
In particular, the extended Liu technique requires to add to the entropy inequality a linear combination of the field equations and of the spatial gradients of the latter (in the following they are called extended equations), up to the order of the gradients entering the state space. The coefficients of this linear combination are the Lagrange multipliers and depend upon the state variables only. Thus,  the number of independent constraints to be taken into account is never less than that of the unknown elements entering the constitutive equations as independent state variables.  

In this paper,  we start discussing the Liu extended procedure from an abstract mathematical point of view, 
and then apply the results to some
physical instances of continua with non-local constitutive equations. In fact, we consider a system of first order balance laws in one space dimension sufficiently general to contain the equations governing the thermodynamical processes occurring in a continuous medium. We assume that this system involves some functions whose constitutive equations are allowed to be non-local, \emph{i.e.}, we admit the possibility that the \emph{state space} includes the gradients of (not necessarily all) the field variables up to the order $r\ge 1$. The compatibility of the constitutive equations with an entropy-like inequality  is then discussed. Once we expand the derivatives in the
entropy-like inequality and impose as constraints the balance equations and some of their gradient extensions,
a set of sufficient conditions, such that the entropy-like inequality is not violated, are  derived. 
Furthermore, the results are applied to two physical instances of continua: 
the first example is concerned with a fluid with a scalar internal variable and constitutive equations with first order 
non-localities, the second one to a fluid whose  state space includes also the second order spatial derivative of the mass density (Korteweg fluid \cite{Korteweg}).

The plan of the paper is the following. In Section~\ref{sec:balance}, we define  mathematically the problem, and introduce the notation that
will be used throughout the paper. In Section~\ref{sec:liu}, we discuss the extended Liu procedure and state the theorem providing the sufficient conditions in order the entropy-like inequality be satisfied for all thermodynamical processes. In Section~\ref{sec:applications},
we give two non-trivial applications of the procedure in meaningful physical situations, and solve the thermodynamical conditions providing explicitly a solution for the constitutive equations. Finally, Section~\ref{sec:conclusions} contains some concluding remarks.

\section{Balance equations for a continuous medium}
\label{sec:balance}
Physical laws describing the mechanical as well as the thermodynamical 
properties of continuous media are usually expressed in terms of balances of some physical quantities (mass, linear and angular momentum, energy, etc.). Here, for simplicity, we consider the case of one-dimensional continuous media and postpone to a forthcoming paper the multi-dimensional case. 

Let 
\[
\mathbf{u}\equiv(u_1(t,x),\ldots,u_n(t,x))
\] 
be the vector of $n$ field variables, depending on time $t$ and space $x$, describing a one-dimensional continuum. 

It is well known that the general governing equations of a continuum are underdetermined since they involve some constitutive functions that specify the particular continuum we are dealing with. 
The various constitutive quantities (for instance, the Cauchy stress tensor or the heat flux) depend on the so called \emph{state variables}: the state variables include (some of) the field variables when a local constitutive theory is adopted, or,  when non-local constitutive theories are considered, also the spatial derivatives up to a finite order $r$ of some field variables. In the following, we shall be concerned with
a non-local constitutive theory. 
In fact, in our framework, the state variables will be the elements of the set 
\[
\displaystyle \mathcal{Z}=\bigcup_{k=0}^r \mathcal{Z}^{(k)},
\]
where
\[
\begin{aligned}
&\mathcal{Z}^{(0)}\subseteq \{u_1,\ldots,u_n\}\equiv \mathcal{U}^{(0)},\\
&\mathcal{Z}^{(k)}\subseteq \left\{\frac{\partial^k u_1}{\partial x^k},\ldots,\frac{\partial^k u_n}{\partial x^k}\right\}\equiv \mathcal{U}^{(k)}, \qquad k=1,\ldots,r,
\end{aligned}
\]
where $r\ge 1$.
Finally, let us denote with $\mathbf{z}$ the vector whose $N$ components belong to the set $\mathcal{Z}$ of state variables, and $\mathbf{z}^\star$ the vector whose $N^\star$ components belong to the set $\mathcal{Z}^\star=\mathcal{U}^{(0)}\bigcup \mathcal{Z}$.

In local form, the thermomechanical description of a continuum leads to consider a system of partial differential equations having the form
\begin{equation}
\label{generalbalance}
\mathcal{E}_i\equiv
\frac{D\Phi_i(\mathbf{u})}{D t}+\frac{D \left(\Psi_i(\mathbf{u})+\chi_i(\mathbf{z}^\star)\right)}{D x}-\Gamma_i(\mathbf{z}^\star)=0, \qquad i=1,\ldots, n,
\end{equation}
where the differential operators $D/Dt$ and $D/Dx$, acting on a composite function $F$ depending on $t$ and $x$ through some quantities $w_1(t,x),\ldots, w_n(t,x)$
(in the paper, these variables are the field variables or the elements of the state space), by chain rule, are defined as follows
\[
\frac{D F}{Dt}=\sum_{j=1}^n\frac{\partial F}{\partial w_j}\frac{\partial w_j}{\partial t},\qquad
\frac{D F}{Dx}=\sum_{j=1}^n\frac{\partial F}{\partial w_j}\frac{\partial w_j}{\partial x}.
\]
These operators do not correspond to total derivatives, and we introduce them in order to avoid confusion when stating Theorem 1, see below.
In the equations (\ref{generalbalance}),
\begin{itemize}
\item $\Phi_i(\mathbf{u})$ are some densities, depending at most on the field variables;
\item the fluxes are split as sums of functions $\Psi_i(\mathbf{u})$ depending at most on the field variables, and 
$\chi_i(\mathbf{z}^\star)$  depending at most on the field variables and the state variables;
\item $\Gamma_i(\mathbf{z}^\star)$ are the production terms depending at most on the field variables and the state variables. 
\end{itemize}

It is well known that in general the constitutive functions have to satisfy some universal principles  
(invariance with respect to rigid motions, time translation, 
scale changes of fundamental quantities, Galilei or Lorentz transformations, etc.). In such a framework, general representation theorems for isotropic scalar, vectorial or tensorial constitutive equations have to be taken into account \cite{Wang1,Wang2,Smith1,Smith2,Smith3}. 
Additional constraints are imposed by the second law of thermodynamics, which requires that the admissible processes must be such that the entropy production is non-negative.

Therefore, in our general framework, we have to exploit the compatibility of constitutive relations with an entropy-like inequality, assumed in the form
\begin{equation}\label{entropyineq1}
\frac{D s(\mathbf{z})}{D t}+\frac{D(v s(\mathbf{z}) +J_s(\mathbf{z}))}{D x}\ge 0,
\end{equation}
where  $s$ and $J_s$, which are functions depending on the state variables, represent the entropy and the entropy flux, respectively, and $v$ the velocity. In the case where the velocity does not appear among the field variables (for instance, in the case of a model of rigid heat conductors), 
the entropy-like inequality is assumed to be
\begin{equation}\label{entropyineq2}
\frac{D s(\mathbf{z})}{D t}+\frac{D J_s(\mathbf{z})}{D x}\ge 0.
\end{equation}

In what follows, we will use the entropy-like inequality in the form \eqref{entropyineq1};   
in the cases where the velocity does not belong to the field variables, the corresponding results are obtained simply by setting $v=0$.

From the analysis of the entropy inequality, some restrictions on the form of constitutive 
equations can be obtained.
Classically, the  restrictions placed by the entropy principle on the constitutive functions are found by using the Coleman-Noll  procedure \cite{Coleman-Noll,Coleman-Mizel}, or the Liu  one \cite{Liu}.
Both procedures have to be extended in order to manage non-local constitutive equations. 
The entropy principle imposes that the inequality \eqref{entropyineq1} must be satisfied for arbitrary 
thermodynamic processes \cite{CJRV-2014,JCL-2010}. To find a set of conditions which are at least sufficient for the 
fulfilment of such a constraint, we apply an extended Liu procedure recently developed in a series of papers 
\cite{Cimmelli-2007,CST-JMP-2009,CST-PRSA-2010,COT-JMP-2011}, 
incorporating new restrictions consistent with higher order non-local constitutive 
theories. In fact, in order to exploit the second law, we use as constraints both the balance equations for the unknown fields and their extended equations up to the order of the 
derivatives entering the state space.

Simple mathematical considerations may clarify the necessity of imposing as additional constraints in the 
entropy inequality the gradients of the balance equations when dealing with non-local constitutive equations. 

The thermodynamic processes are solutions of the balance equations, and, if these solutions are smooth enough, 
are trivially solutions of their differential consequences  (see also \cite{Rogolino-Cimmelli-2019}). 
Since the entropy inequality \eqref{entropyineq1} has to be satisfied in arbitrary smooth processes,  then it is 
natural, from a mathematical point of view, to use the differential consequences of the equations governing those 
processes as constraints for such an inequality. 
Next Section will be devoted to the description of the extended Liu procedure in this general framework.
 
\section{Extended Liu procedure}
\label{sec:liu}
In this Section, we introduce a general scheme in order to apply the extended Liu technique in the 
case of $r$-th order ($r\ge 1$) non-local constitutive equations.

We consider this rather general case essentially for two reasons. The first reason is to have a unified framework good enough to be applied to different models with non-local constitutive equations of arbitrary order. The second one is of computational nature. In fact, in dealing with applied problems, it is relevant both the derivation of the thermodynamic restrictions arising from the entropy inequality, and, whenever this is possible, a more or less explicit 
characterization of the constitutive equations. Since the thermodynamic restrictions may have  a lengthy expression, it is convenient to use a computer algebra system for their possible solution. In order to be able to have a flexible computer algebra package that can be used in many different cases, a general approach reveals useful if not necessary. In fact, we developed some general routines in the CAS Reduce \cite{Reduce} that implement the algorithm at the core of extended Liu approach. 

First of all, we need to compute the spatial derivatives of the fundamental balance equations. 
From  the system \eqref{generalbalance}, developing the first order time and space derivatives, one has:
\begin{equation}
\mathcal{E}_i\equiv \frac{\partial \Phi_i}{\partial u_j}\frac{\partial u_j}{\partial t}+\frac{\partial \Psi_i}{\partial u_j}\frac{\partial u_j}{\partial x}+
\frac{\partial \chi_i}{\partial z^\star_{\alpha}}\frac{\partial z^\star_\alpha}{\partial x}-\Gamma_i=0,
\end{equation}
where the Einstein summation convention over repeated indices is used. 
In order to write in general the $m$-th order spatial derivative of the balance laws, let us recall 
\cite{Mishkov} a formula giving an expression for the $m$-th derivative of a composite function when the argument is a vector with an arbitrary number of components. This formula is a generalization of the well known Fa\`a di Bruno's formula \cite{FaadiBruno,Roman}.

The following theorem is no more than a simple rewriting of the main result contained in \cite{Mishkov}.
\begin{theorem}
Let $\mathbf{w}=(w_1(t,x),\ldots,w_s(t,x))$ be a vector, and $F(\mathbf{w}(t,x))$ a composite function for which all the needed derivatives are defined, then
\begin{equation}
\label{faabruno}
\begin{aligned}
&\frac{D^m F(\mathbf{w}(t,x))}{D x^m}=
\sum_{J_0}\sum_{J_1}\cdots\sum_{J_m}
\frac{m!}{\prod_{i=1}^m(i!)^{k_i}\prod_{i=1}^m\prod_{j=1}^s q_{ij}!}\times\\
&\qquad\times \frac{\partial^k F}{\partial w_1^{p_1}\partial w_2^{p_2}\cdots\partial w_s^{p_s}}
\prod_{i=1}^m \left(\frac{\partial^i w_1}{\partial x^i}\right)^{q_{i1}}
\left(\frac{\partial^i w_2}{\partial x^i}\right)^{q_{i2}}\cdots 
\left(\frac{\partial^i w_s}{\partial x^i}\right)^{q_{is}},
\end{aligned}
\end{equation}
where the various sums are over all nonnegative integer solutions of the Diophantine equations
\begin{equation*}
\begin{aligned}
&\sum_{J_0} \rightarrow k_1+2k_2+\ldots +mk_m = m,\\
&\sum_{J_1} \rightarrow q_{11}+q_{12}+\ldots +q_{1s} = k_1,\\
&\sum_{J_2} \rightarrow q_{21}+q_{22}+\ldots +q_{2s} = k_2,\\
&\ldots\\
&\sum_{J_m} \rightarrow q_{m1}+q_{m2}+\ldots +q_{ms} = k_m,
\end{aligned}
\end{equation*}
and it is
\begin{equation*}
\begin{aligned}
&p_j=q_{1j}+q_{2j}+\ldots+q_{mj}, \qquad j=1,\ldots,s,\\
&k=p_1+p_2+\ldots+p_s=k_1+k_2+\ldots+k_m. \qquad \square
\end{aligned}
\end{equation*}
\end{theorem}

By introducing the $m$-th order $(m=1,\ldots,r)$ spatial derivatives of the balance laws, we get
\begin{equation}
\label{estese}
\begin{aligned}
&\frac{D^m \mathcal{E}_i}{D x^m}\equiv\sum_{h=0}^k\binom{m}{h}\left[\frac{D^{h}}{D x^{h}}\left(\frac{\partial \Phi_i(\mathbf{u})}{\partial u_j}\right)\frac{\partial^{m-h+1}u_j}{\partial t \partial x^{m-h}} \right.\\
&\left.+\frac{D^{h}}{D x^{h}}\left(\frac{\partial \Psi_i(\mathbf{u})}{\partial u_j}\right)\frac{\partial^{m-h+1}u_j}{\partial x^{m-h+1}}+\frac{D^h}{D x^h}
\left(\frac{\partial \chi_i(\mathbf{z}^\star)}{\partial z^\star_\alpha}\right)\frac{\partial^{m-h+1}z^\star_\alpha}{\partial x^{m-h+1}}\right]\\
&-\frac{D^m \Gamma_i(\mathbf{z}^\star)}{D x^{m}}=0.
\end{aligned}
\end{equation}
In order to take into account in the entropy inequality the restrictions determined by the field equations and their spatial derivatives, let us introduce the Lagrange multipliers $\Lambda^{(k)}_i$
$(i=1,\ldots,n,\; k=0,\ldots,r)$ associated to 
$\displaystyle\frac{D^k\mathcal{E}_i}{D x^k}$. Therefore, the entropy inequality \eqref{entropyineq1} becomes
\begin{equation}\label{dis}
\frac{\partial s(\mathbf{z})}{\partial z_\alpha}\frac{\partial z_\alpha}{\partial t}+v\frac{\partial s(\mathbf{z})}{\partial z_\alpha}\frac{\partial z_\alpha}{\partial x}+s(\mathbf{z})\frac{\partial v}{\partial x}+\frac{\partial J_s(\mathbf{z})}{\partial z_\alpha}\frac{\partial z_\alpha}{\partial x}-\sum_{i=1}^n\sum_{k=0}^r
\Lambda_i^{(k)}\frac{D^k \mathcal{E}_i}{D x^k}\geq 0.
\end{equation}

By expanding the derivatives in \eqref{dis}, and using formulae \eqref{faabruno} and \eqref{estese}, a straightforward though tedious computation provides a very long expression that turns out to be a polynomial in some derivatives of field variables not belonging to the state space with coefficients depending at most on the field and state variables. This polynomial must be non-negative! Once \eqref{dis} has been expanded, we may distinguish the derivatives of field variables therein appearing,
and not entering the state space, in two different classes:
\begin{itemize}
\item \emph{highest derivatives}: time derivatives of the field variables, time derivatives of the spatial derivatives (up to the order $r$) of the field variables, and spatial derivatives of highest order: it is easily ascertained that these highest derivatives appear linearly with coefficients depending on the field and state variables;
\item \emph{higher derivatives}: spatial derivatives whose order is not maximal but higher than that of the derivatives entering the state space: it is easily recognized that these higher derivatives appear in powers of degree up to $r+1$, with coefficients depending on the field and state variables.
\end{itemize}

Let us define the following sets:
\[
\begin{aligned}
&\widehat{\mathcal{Z}}^{(k)}=\left\{ w \in \mathcal{Z}^{(k)} 
\;:\; \frac{\partial^h w}{\partial x^h}\notin \mathcal{Z},\; h=1,\ldots,r-k\right\},\quad k=0,\ldots,r-1,\\
&\widehat{\mathcal{Z}}^{(r)}\equiv \mathcal{Z}^{(r)},
\end{aligned}
\]
and
\[
\begin{aligned}
&\widehat{\mathcal{Z}}=\bigcup_{k=0}^r \widehat{\mathcal{Z}}^{(k)}.
\end{aligned}
\]
The highest derivatives are the time derivatives of the field variables and of their spatial derivatives up to the order $r$, together with the $(r+1)$th order spatial derivatives of the elements belonging to the set $\widehat{\mathcal{Z}}$.

Therefore, denoting with  $\boldsymbol\zeta$ the vector whose components $\zeta_i$ are the highest derivatives, and with 
$\boldsymbol\eta$ the vector whose components $\eta_j$ are the higher derivatives, 
the entropy inequality \eqref{dis} can be cast in the following compact form:
\begin{equation}
\begin{aligned}
A_i(\mathbf{z}^\star) \zeta_i&+B^{(r+1)}_{j_1\ldots j_{r+1}}(\mathbf{z}^\star)\eta_{j_1}\cdots \eta_{j_{r+1}}+
B^{(r)}_{j_1\ldots j_{r}}(\mathbf{z}^\star)\eta_{j_1}\cdots \eta_{j_{r}}\\
&+\ldots+B^{(2)}_{j_1j_{2}}(\mathbf{z}^\star)\eta_{j_1}\eta_{j_{2}}+B^{(1)}_{j_1}(\mathbf{z}^\star)\eta_{j_1}+B^{(0)}(\mathbf{z}^\star)\ge 0,
\end{aligned}
\end{equation}
where the coefficients $A_i$, $B^{(r+1)}_{j_1\ldots j_{r+1}}$, $B^{(r)}_{j_1\ldots j_{r}}$,\ldots,
$B^{(2)}_{j_1j_{2}}$,  $B^{(1)}_{j_1}$ and $B^{(0)}$ may depend upon the field variables and the elements entering the state space.
This inequality must be satisfied for every thermodynamical process. 

First, let us observe that nothing prevents to have a thermodynamic process where $B^{(0)}=0$.
Moreover, since we used in the entropy inequality all the constraints imposed by
the field equations together with their spatial derivatives, the highest and higher derivatives may assume arbitrary values. 
Consequently, we may give a set of conditions that are sufficient in order the inequality \eqref{dis} be fulfilled for every thermodynamical process. These sufficient conditions provide constraints on the constitutive equations.

\begin{theorem}\label{theorem2}
Let $\boldsymbol\zeta=(\zeta_1,\ldots,\zeta_p)$ be the vector of highest derivatives, and $\boldsymbol\eta=(\eta_1,\ldots,\eta_q)$ the vector of higher derivatives.  
Let 
\begin{itemize}
\item $A_i(\mathbf{\mathbf{z}^\star})$ are $p$ functions of $\mathbf{z}^\star$;
\item $B^{(k)}_{j_1\ldots j_{k}}(\mathbf{z}^\star)$, with $k=1,\ldots,r+1$, are $\binom{q-k+1}{k}$ functions of $\mathbf{z}^\star$; 
\item $B^{(0)}(\mathbf{z}^\star)$ is a function of $\mathbf{z}^\star$.
\end{itemize}
The inequality
\begin{equation}
\begin{aligned}
A_i(\mathbf{z}^\star) \zeta_i&+B^{(r+1)}_{j_1\ldots j_{r+1}}(\mathbf{z}^\star)\eta_{j_1}\cdots \eta_{j_{r+1}}+
B^{(r)}_{j_1\ldots j_{r}}(\mathbf{z}^\star)\eta_{j_1}\cdots \eta_{j_{r}}\\
&+\ldots+B^{(2)}_{j_1j_{2}}(\mathbf{z}^\star)\eta_{j_1}\eta_{j_{2}}+B^{(1)}_{j_1}(\mathbf{z}^\star)\eta_{j_1}+B^{(0)}(\mathbf{z}^\star)\ge 0,
\end{aligned}
\end{equation}
holds for arbitrary vectors $\boldsymbol\zeta$ and $\boldsymbol\eta$ if 
\begin{enumerate}
\item $A_i=0$;
\item $B^{(2k-1)}_{j_1\ldots j_{2k-1}}=0$, $k=1,\ldots,\lfloor{\frac{r+2}{2}}\rfloor$\footnote{$\lfloor{x}\rfloor$ denotes the greatest integer less than $x$.};
\item $B^{(2k)}_{j_1\ldots j_{2k}}\eta_{j_1}\cdots \eta_{j_{2k}}$, $k=1,\ldots,\lfloor{\frac{r+1}{2}}\rfloor$,
 nonnegative for all $\boldsymbol\eta$;
\item $B^{(0)}\ge 0$. $\qquad \square$
\end{enumerate}

\end{theorem}

Due to Theorem~\ref{theorem2}, by imposing that the coefficients of $u_{j,t}$, $u_{j,t x}$, \ldots, $u_{j,t,\underbrace{x\ldots x}_{r}}$, where the indices ${(\cdot)}_{,t}$ and ${(\cdot)}_{,x}$ denote the
partial derivatives with respect to time and space, respectively,
are vanishing, we obtain:
\begin{equation}
\begin{aligned}
&\frac{\partial s}{\partial u_j}-\sum_{i=1}^n\sum_{k=0}^r\Lambda_i^{(k)}\frac{D^k}{Dx^k}\left(\frac{\partial\Phi_i}{\partial u_j}\right)=0,\\
&\frac{\partial s}{\partial u_{j,x}}-\sum_{i=1}^n\sum_{k=1}^r \binom{k}{k-1}\Lambda_i^{(k)}\frac{D^{k-1}}{Dx^{k-1}}\left(\frac{\partial\Phi_i}{\partial u_j}\right)=0,\\
&\frac{\partial s}{\partial u_{j,xx}}-\sum_{i=1}^n\sum_{k=2}^r \binom{k}{k-2}\Lambda_i^{(k)}\frac{D^{k-2}}{Dx^{k-2}}\left(\frac{\partial\Phi_i}{\partial u_j}\right)=0,\\
&\ldots\\
&\frac{\partial s}{\partial u_{j,\underbrace{x\ldots x}_{r}}}-\sum_{i=1}^n\Lambda_i^{(r)}\frac{\partial\Phi_i}{\partial u_j}=0,
\end{aligned}
\end{equation}
that allow us the determination of the Lagrange multipliers. The coefficients of the remaining highest derivatives, 
\begin{equation}
\sum_{i=1}^n\Lambda_i^{(r)}\left(\frac{\partial\Psi_i}{\partial \widehat{z}_\alpha}+\frac{\partial\chi_i}{\partial \widehat{z}_\alpha}\right)=0,\qquad \widehat{z}_\alpha\in\widehat{\mathcal{Z}},
\end{equation}
provide conditions to be used together with the constraints coming from 
the arbitrariness of the higher derivatives to restrict the constitutive equations. After all these restrictions have been derived, the residual entropy inequality, say
\begin{equation} 
B^{(0)}\ge 0,
\end{equation}
remains providing further constraints.

It is evident that the general restrictions here derived can not be discussed
from a physical point of view, but they are essential in writing the computer algebra program that almost automatically computes the restrictions placed by second law.

In the next Section, we provide some examples of physical interest where the procedure here described can be applied, and we discuss the physical meaning of the results.

\section{Applications}
\label{sec:applications}
Here, we consider two physical examples of fluids whose constitutive equations involve first or second order non-localities, \emph{i.e.}, special instances of higher grade fluids. 
In modern terminology, a fluid is said to be of grade $(r+1)$ 
\cite{Dunn-Serrin,Truesdell_Noll,Dunn-Rajagopal,Gouin2019} if the constitutive quantities are allowed to depend on gradients of order  $r$.  In recent years, these higher grade 
fluids have been employed, for instance, to model capillarity effects 
\cite{Gouin1985a,Gouin1985b}, or to analyze the structure of liquid-vapor phase transitions 
under both static
\cite{Aifantis-Serrin-1,Aifantis-Serrin-2} and dynamic \cite{Slemrod-1,Slemrod-2} conditions. 

As observed in \cite{Dunn-Serrin,Dunn}, these fluids are, in general, incompatible with the restrictions placed by the second law of thermodynamics. In order to find a remedy to such an incompatibility, these authors proposed a generalization of the classical local balance of energy by postulating the existence of a rate of supply of mechanical energy, the so called interstitial working; in such a framework, the entropy flux has the classical form as the ratio between the heat flux and the absolute temperature. Nevertheless, the same authors \cite{Dunn-Serrin} remarked that the interstitial working can be removed but at the cost of introducing an entropy extra-flux \cite{Muller} in order to satisfy the second law of thermodynamics. 

In the applications we consider below, the assumptions we make consist in taking the local energy balance in the classical form without including any extra-term; moreover, we write the entropy inequality without specializing the form of the entropy flux: as will be seen, the expression of entropy flux will arise as a consequence of the extended procedure when solving the constraints placed by the second law.

\subsection{Fluid of grade 2 with a scalar internal variable}
Let us consider a fluid of grade 2 whose description involves, in addition to the basic fields of mass density, velocity and internal energy, an internal variable. The latter may describe an additional internal degree of freedom of the material, for instance representative of a suitable scalar microstructure \cite{Capriz,OS-2008} or another extensive property.

The governing equations we consider read
\begin{equation}
\label{fluid-grade-2}
\begin{aligned}
&\mathcal{E}_1\equiv\frac{D\rho}{D t} + \frac{D (\rho v)}{D x}=0, \\
&\mathcal{E}_2\equiv\frac{D(\rho v)}{D t} + \frac{D (\rho v^2  - T)}{D x}=0,\\
&\mathcal{E}_3\equiv\frac{D}{D t}\left(\rho\varepsilon+\rho \frac{v^2}{2}\right) +\frac{D }{D x}\left(\rho v\varepsilon+\rho \frac{v^3}{2}-Tv+ q\right)=0,\\
&\mathcal{E}_4\equiv\frac{D(\rho\gamma)}{Dt}+\frac{D(\rho v\gamma+\phi)}{Dx}=0,
\end{aligned}
\end{equation}
where $\rho$ is the mass density, $v$ the velocity, $\varepsilon$ the internal energy per unit mass, and $\gamma$ an internal state variable;
moreover, the Cauchy stress $T$, the heat flux $q$, and the flux $\phi$  of internal variable must be assigned by 
means of suitable constitutive equations such that for every admissible process the entropy inequality 
\begin{equation}
\rho\left(\frac{Ds}{D t}+v\frac{Ds}{Dx}\right)+ \frac{DJ_s}{Dx} \ge 0
\end{equation}
be satisfied, being $s$ and $J_s$ (to be assigned as constitutive quantities) the specific entropy and  the entropy flux, respectively.

We assume the state space spanned by
\begin{equation}
\mathcal{Z}=\{\rho,\varepsilon,\gamma,\rho_{,x},v_{,x},\varepsilon_{,x},\gamma_{,x}\}.
\end{equation}

As shown in the previous Section, the exploitation of second law of thermodynamics is here performed by taking into 
account the constraints 
imposed on the thermodynamic processes by the balance equations and their first order extensions; these 
constraints are imposed by introducing some  Lagrange multipliers.
Therefore, the entropy inequality becomes
\begin{equation}
\label{entropyconstrained}
\begin{aligned}
&\rho\left(\frac{Ds}{D t}+v\frac{Ds}{Dx}\right)+ \frac{DJ_s}{Dx} \\
&\quad- \Lambda^{(0)}_1 \mathcal{E}_1- \Lambda^{(0)}_2 \mathcal{E}_2- \Lambda^{(0)}_3 \mathcal{E}_3
- \Lambda^{(0)}_4 \mathcal{E}_4\\
&\quad-\Lambda^{(1)}_1\frac{D\mathcal{E}_1}{Dx}-\Lambda^{(1)}_2\frac{D\mathcal{E}_2}{Dx}
-\Lambda^{(1)}_3\frac{D\mathcal{E}_3}{Dx}-\Lambda^{(1)}_4\frac{D\mathcal{E}_4}{Dx} \geq 0.
\end{aligned}
\end{equation}
For the sake of clarity, we present the details of the computation we are required to do in applying the extended Liu procedure. 

Expanding the derivatives in the entropy inequality \eqref{entropyconstrained}, we obtain a very long expression, say
\begin{align*}
&\left(\rho\frac{\partial s}{\partial\rho}-\Lambda^{(0)}_1\right)\rho_{,t}-\left(\rho\Lambda^{(0)}_2+\rho_{,x}\Lambda^{(1)}_2\right)v_{,t}+\left(\rho\frac{\partial s}{\partial\varepsilon}-\rho\Lambda^{(0)}_3-\rho_{,x}\Lambda^{(1)}_3\right)\varepsilon_{,t}\allowdisplaybreaks\\
&\quad +\left(\rho\frac{\partial s}{\partial\gamma}-\rho\Lambda^{(0)}_4-\rho_{,x}\Lambda^{(1)}_4\right)\gamma_{,t}
+\left(\rho\frac{\partial s}{\partial\rho_{,x}}-\Lambda^{(1)}_1\right)\rho_{,tx}\allowdisplaybreaks\\
&\quad+\rho\left(\frac{\partial s}{\partial v_{,x}}-\Lambda^{(1)}_2\right)v_{,tx}
+\rho\left(\frac{\partial s}{\partial \varepsilon_{,x}}-\Lambda^{(1)}_3\right)\varepsilon_{,tx}
+\rho\left(\frac{\partial s}{\partial \gamma_{,x}}-\Lambda^{(1)}_4\right)\gamma_{,tx}\allowdisplaybreaks\\
&\quad+\left(\frac{\partial T}{\partial\rho_{,x}}\Lambda^{(1)}_2- \frac{\partial q}{\partial\rho_{,x}}\Lambda^{(1)}_3 
- \frac{\partial\phi}{\partial\rho_{,x}}\Lambda^{(1)}_4 \right)\rho_{,xxx}\allowdisplaybreaks\\
&\quad+\left(\frac{\partial T}{\partial v_{,x}}\Lambda^{(1)}_2- \frac{\partial q}{\partial v_{,x}}\Lambda^{(1)}_3 
- \frac{\partial\phi}{\partial v_{,x}}\Lambda^{(1)}_4 \right)v_{,xxx}\allowdisplaybreaks\\
&\quad +\left(\frac{\partial T}{\partial \varepsilon_{,x}}\Lambda^{(1)}_2 - \frac{\partial q}{\partial \varepsilon_{,x}}\Lambda^{(1)}_3 
- \frac{\partial\phi}{\partial \varepsilon_{,x}}\Lambda^{(1)}_4\right)\varepsilon_{,xxx}\allowdisplaybreaks\\
&\quad +\left(\frac{\partial T}{\partial \gamma_{,x}}\Lambda^{(1)}_2- \frac{\partial q}{\partial \gamma_{,x}}\Lambda^{(1)}_3 
- \frac{\partial\phi}{\partial \gamma_{,x}}\Lambda^{(1)}_4 \right)\gamma_{,xxx}\allowdisplaybreaks\\
&\quad+\left(\frac{\partial ^{2}T}{\partial \rho_{,x}^{2}}  \Lambda^{(1)}_2 
- \frac{\partial ^{2}q}{\partial \rho_{,x}^{2}}  \Lambda^{(1)}_3
-\frac{\partial ^{2}\phi }{\partial \rho_{,x}^{2}}  \Lambda^{(1)}_4 
\right)  \rho_{,xx}^{2}\allowdisplaybreaks\\
&\quad+2 \left(\frac{\partial ^{2}T}{\partial \rho_{,x}\partial v_{,x}}  \Lambda^{(1)}_2   
- \frac{\partial ^{2}q}{\partial \rho_{,x}\partial v_{,x}}  \Lambda^{(1)}_3
-\frac{\partial ^{2}\phi }{\partial \rho_{,x}\partial v_{,x}}  \Lambda^{(1)}_4\right)  \rho_{,xx}  v_{,xx}\allowdisplaybreaks\\  
&\quad+2\left( \frac{\partial ^{2}T}{\partial \varepsilon _{,x}\partial \rho_{,x}}    \Lambda^{(1)}_2 
- \frac{\partial ^{2}q}{\partial \varepsilon _{,x}\partial \rho_{,x}} \Lambda^{(1)}_3
- \frac{\partial ^{2}\phi }{\partial \varepsilon _{,x}\partial \rho_{,x}}    \Lambda^{(1)}_4  \right) \rho_{,xx} \varepsilon _{,xx}    \allowdisplaybreaks\\
&\quad+2 \left( \frac{\partial ^{2}T}{\partial \gamma_{,x}\partial \rho_{,x}} \Lambda^{(1)}_2 
-\frac{\partial ^{2}q}{\partial \gamma_{,x}\partial \rho_{,x}} \Lambda^{(1)}_3 
- \frac{\partial ^{2}\phi }{\partial \gamma_{,x}\partial \rho_{,x}} \Lambda^{(1)}_4\right)  \rho_{,xx}\gamma_{,xx}  \allowdisplaybreaks\\
&\quad+\left(\frac{\partial ^{2}T}{\partial v_{,x}^{2}}  \Lambda^{(1)}_2 
- \frac{\partial ^{2}q}{\partial v_{,x}^{2}}  \Lambda^{(1)}_3 
-\frac{\partial ^{2}\phi }{\partial v_{,x}^{2}}  \Lambda^{(1)}_4\right)  v_{,xx}^{2}\allowdisplaybreaks\\
&\quad+2\left( \frac{\partial ^{2}T}{\partial \varepsilon _{,x}\partial v_{,x}}  \Lambda^{(1)}_2 
- \frac{\partial ^{2}q}{\partial \varepsilon _{,x}\partial v_{,x}} \Lambda^{(1)}_3 
- \frac{\partial ^{2}\phi }{\partial \varepsilon _{,x}\partial v_{,x}} \Lambda^{(1)}_4\right) v_{,xx} \varepsilon _{,xx}   \allowdisplaybreaks\\
&\quad+2\left(  \frac{\partial ^{2}T}{\partial \gamma_{,x}\partial v_{,x}}  \Lambda^{(1)}_2  
- \frac{\partial ^{2}q}{\partial \gamma_{,x}\partial v_{,x}}   \Lambda^{(1)}_3  
- \frac{\partial ^{2}\phi }{\partial \gamma_{,x}\partial v_{,x}}\Lambda^{(1)}_4\right)  v_{,xx}\gamma_{,xx} \allowdisplaybreaks\\
&\quad+ \left(\frac{\partial ^{2}T}{\partial \varepsilon _{,x}^{2}} \Lambda^{(1)}_2
- \frac{\partial ^{2}q}{ \partial \varepsilon _{,x}^{2}} \Lambda^{(1)}_3 
-\frac{\partial ^{2}\phi }{\partial \varepsilon _{,x}^{2}} \Lambda^{(1)}_4\right) \varepsilon _{,xx}^{2}  \allowdisplaybreaks\\
&\quad+2  \left(\frac{\partial ^{2}T}{\partial \varepsilon _{,x}\partial \gamma_{,x}}\Lambda^{(1)}_2  
- \frac{\partial ^{2}q}{\partial \varepsilon _{,x}\partial \gamma_{,x}}  \Lambda^{(1)}_3 
- \frac{\partial ^{2}\phi }{\partial \varepsilon _{,x}\partial \gamma_{,x}}\Lambda^{(1)}_4 \right) \varepsilon _{,xx}  \gamma_{,xx}
  \allowdisplaybreaks\\
&\quad+\left(\frac{\partial ^{2}T}{\partial \gamma_{,x}^{2}}   \Lambda^{(1)}_2
- \frac{\partial ^{2}q}{\partial \gamma_{,x}^{2}}   \Lambda^{(1)}_3
-\frac{\partial ^{2}\phi }{\partial \gamma_{,x}^{2}}  \Lambda^{(1)}_4\right) \gamma_{,xx}^{2} \allowdisplaybreaks\\
&\quad+\left(\frac{\partial s}{\partial \rho_{,x}}  \rho  v+\frac{\partial J_s}{\partial \rho_{,x}} +\frac{\partial T}{\partial \rho_{,x}}  \Lambda^{(0)}_2- \frac{\partial q}{\partial \rho_{,x}}  \Lambda^{(0)}_3-\frac{\partial \phi }{\partial \rho_{,x}}  \Lambda^{(0)}_4\right.\allowdisplaybreaks\\
&\qquad-\Lambda^{(1)}_1 v+\Lambda^{(1)}_2\left(\frac{\partial T}{\partial \rho }  +2 \frac{\partial ^{2}T}{\partial \rho \partial \rho_{,x}}  \rho_{,x}  +2 \frac{\partial ^{2}T}{\partial \varepsilon \partial \rho_{,x}}  \varepsilon _{,x}    
+2 \frac{\partial ^{2}T}{\partial \gamma\partial \rho_{,x}}  \gamma_{,x}\right) \allowdisplaybreaks \\
&\qquad +\Lambda^{(1)}_3\left(\frac{\partial T}{\partial \rho_{,x}}  v_{,x}-2 \frac{\partial ^{2}q}{\partial \varepsilon \partial \rho_{,x}}  \varepsilon _{,x}  -2  \frac{\partial ^{2}q}{\partial \gamma\partial \rho_{,x}}  \gamma_{,x}  
-2 \frac{\partial ^{2}q}{\partial \rho \partial \rho_{,x}}   \rho_{,x}  
-\frac{\partial q}{\partial \rho }  \right)    \allowdisplaybreaks\\
&\qquad\left.+\Lambda^{(1)}_4\left(  -2  \frac{\partial ^{2}\phi }{ \partial \varepsilon \partial \rho_{,x}}  \varepsilon _{,x}    
-2\frac{\partial ^{2}\phi }{\partial \gamma\partial \rho_{,x}}  \gamma_{,x}   
-2\frac{\partial ^{2}\phi }{\partial \rho \partial \rho_{,x}}  \rho_{,x} -\frac{\partial \phi }{\partial \rho } \right)
\right) \rho_{,xx} \allowdisplaybreaks\\
&\quad+ \left(\frac{\partial s}{\partial v_{,x}}  \rho   v  
+\frac{\partial J_s}{\partial v_{,x}}  
+\frac{\partial T}{\partial v_{,x}}  \Lambda^{(0)}_2  
- \frac{\partial q}{\partial v_{,x}}  \Lambda^{(0)}_3  
-\frac{\partial \phi }{\partial v_{,x}}  \Lambda^{(0)}_4\right.\allowdisplaybreaks\\
&\qquad+\left(-\Lambda^{(1)}_1  \rho   
-\Lambda^{(1)}_2  \left(\rho   v  
+2 \frac{\partial ^{2}T}{\partial \varepsilon \partial v_{,x}}  \varepsilon _{,x}  
+2 \frac{\partial ^{2}T}{\partial \gamma\partial v_{,x}}  \gamma_{,x}  
+2 \frac{\partial ^{2}T}{\partial \rho \partial v_{,x}}    \rho_{,x}\right)\right.\allowdisplaybreaks\\
&\qquad+\Lambda^{(1)}_3  \left(T +\frac{\partial T}{\partial v_{,x}}   v_{,x}
-2 \frac{\partial ^{2}q}{\partial \rho \partial v_{,x}}   \rho_{,x}  
-2 \frac{\partial ^{2}q}{\partial \varepsilon \partial v_{,x}}  \varepsilon _{,x} 
-2 \frac{\partial ^{2}q}{\partial \gamma\partial v_{,x}}  \gamma_{,x} \right) \allowdisplaybreaks\\
&\qquad\left.+\Lambda^{(1)}_4\left(-2\frac{\partial ^{2}\phi }{\partial \rho \partial v_{,x}}\rho_{,x}
-2  \frac{\partial ^{2}\phi }{ \partial \varepsilon \partial v_{,x}}  \varepsilon _{,x}  
-2\frac{\partial ^{2}\phi }{\partial \gamma\partial v_{,x}}  \gamma_{,x}  \right) \right) v_{,xx}\allowdisplaybreaks\\
&\quad+ \left(\frac{\partial s}{\partial \varepsilon _{,x}}   \rho   v
+\frac{\partial J_s}{\partial \varepsilon _{,x}} 
+\frac{\partial T}{\partial \varepsilon _{,x}}  \Lambda^{(0)}_2
 -\frac{\partial q}{\partial \varepsilon _{,x}}  \Lambda^{(0)}_3
-\frac{\partial \phi }{\partial \varepsilon _{,x}} \Lambda^{(0)}_4\right.\allowdisplaybreaks\\
&\qquad+\Lambda^{(1)}_2\left(2  \frac{\partial ^{2}T}{\partial \varepsilon \partial \varepsilon _{,x}}  \varepsilon _{,x} 
+ \frac{\partial T}{\partial \varepsilon }  
+2 \frac{\partial ^{2}T}{\partial \varepsilon _{,x}\partial \gamma}  \gamma_{,x} 
+2 \frac{\partial ^{2}T}{\partial \varepsilon _{,x}\partial \rho } \rho_{,x}\right)\allowdisplaybreaks\\
&\qquad+\Lambda^{(1)}_3\left(-2\frac{\partial ^{2}q}{\partial \varepsilon \partial \varepsilon _{,x}}  \varepsilon _{,x}  
+\frac{\partial T}{\partial \varepsilon _{,x}}  v_{,x}
-  \rho   v
-\frac{\partial q}{\partial \varepsilon } 
-2 \frac{\partial ^{2}q}{\partial \varepsilon _{,x}\partial \gamma} \gamma_{,x}  
-2 \frac{\partial ^{2}q}{\partial \varepsilon _{,x}\partial \rho } \rho_{,x}\right) \allowdisplaybreaks\\
&\qquad\left.+\Lambda^{(1)}_4\left(-2  \frac{\partial ^{2}\phi }{\partial \varepsilon \partial \varepsilon _{,x}}  \varepsilon _{,x} 
-\frac{\partial \phi }{\partial \varepsilon } 
-2  \frac{\partial ^{2}\phi }{\partial \varepsilon _{,x}\partial \gamma}   \gamma_{,x} 
-2 \frac{\partial ^{2}\phi }{\partial \varepsilon _{,x}\partial \rho } \rho_{,x}\right)\right) \varepsilon _{,xx}   \allowdisplaybreaks\\  
&\quad+ \left(\frac{\partial s}{\partial \gamma_{,x}}  \rho   v
+\frac{\partial J_s}{\partial \gamma_{,x}}  
+\frac{\partial T}{\partial \gamma_{,x}}  \Lambda^{(0)}_2
- \frac{\partial q}{\partial \gamma_{,x}}  \Lambda^{(0)}_3
-\frac{\partial \phi }{\partial \gamma_{,x}}  \Lambda^{(0)}_4\right.\allowdisplaybreaks\\
&\qquad+\Lambda^{(1)}_2\left(2 \frac{\partial ^{2}T}{\partial \varepsilon \partial \gamma_{,x}}  \varepsilon _{,x} 
+2  \frac{\partial ^{2}T}{\partial \gamma\partial \gamma_{,x}}  \gamma_{,x} 
+\frac{\partial T}{ \partial \gamma}  
+2  \frac{\partial ^{2}T}{\partial \gamma_{,x}\partial \rho }  \rho_{,x}\right)  \allowdisplaybreaks\\
&\qquad+\Lambda^{(1)}_3\left(-2 \frac{\partial ^{2}q}{\partial \varepsilon \partial \gamma_{,x}}  \varepsilon _{,x}  
-2 \frac{\partial ^{2}q}{\partial \gamma\partial \gamma_{,x}}  \gamma_{,x}  
-\frac{\partial q}{\partial \gamma}  
-2  \frac{\partial ^{2}q}{\partial \gamma_{,x}\partial \rho }  \rho_{,x}
+\frac{\partial T}{\partial \gamma_{,x}} v_{,x}\right)\allowdisplaybreaks\\
&\qquad\left.+\Lambda^{(1)}_4 \left(-2  \frac{\partial ^{2}\phi }{\partial \varepsilon \partial \gamma_{,x}}  \varepsilon _{,x} 
-2\frac{\partial ^{2}\phi }{\partial \gamma\partial \gamma_{,x}}  \gamma_{,x}  
-\frac{\partial \phi }{\partial \gamma} 
-2 \frac{\partial ^{2}\phi }{\partial \gamma_{,x}\partial \rho }  \rho_{,x}
-\rho  v\right)\right)\gamma_{,xx}\allowdisplaybreaks\\
& \rho v\left(\frac{\partial s}{\partial \rho }\rho_{,x} 
+ \frac{\partial s}{\partial \varepsilon }  \varepsilon _{,x} 
+ \frac{\partial s}{\partial \gamma}  \gamma_{,x}\right)
+\frac{\partial J_s}{\partial \rho }  \rho_{,x}
+\frac{\partial J_s}{ \partial \varepsilon }  \varepsilon _{,x}
+\frac{\partial J_s}{ \partial \gamma}  \gamma_{,x}\allowdisplaybreaks\\
&\quad-\Lambda^{(0)}_1  \left(v\rho_{,x}+\rho   v_{,x}\right)
-\Lambda^{(0)}_2 \left(\rho   v  v_{,x}-\frac{\partial T}{\partial \rho } \rho_{,x}
- \frac{\partial T}{\partial \varepsilon }  \varepsilon _{,x} 
-\frac{\partial T}{ \partial \gamma}  \gamma_{,x} \right)\allowdisplaybreaks\\
&\quad-\Lambda^{(0)}_3\left(\rho   v\varepsilon _{,x}  -T  v_{,x}+ \frac{\partial q}{\partial \rho }  \rho_{,x}
+\frac{\partial q}{ \partial \varepsilon }  \varepsilon _{,x}  
+\frac{\partial q}{\partial \gamma}  \gamma_{,x}\right)\allowdisplaybreaks\\
&\quad-\Lambda^{(0)}_4 \left(\rho v\gamma_{,x} +\frac{\partial \phi }{\partial \rho }   \rho_{,x} 
+\frac{\partial \phi }{\partial \varepsilon }  \varepsilon _{,x}  
+\frac{\partial \phi }{\partial \gamma}  \gamma_{,x}  \right)
-2  \Lambda^{(1)}_1  \rho_{,x}  v_{,x}\allowdisplaybreaks\\
&\quad-\Lambda^{(1)}_2\left(v\rho_{,x} v_{,x}+\rho   v_{,x}^{2}
-\frac{\partial ^{2}T}{\partial \rho ^{2}}   \rho_{,x}^{2}
-2 \frac{\partial ^{2}T}{ \partial \rho \partial \varepsilon} \rho_{,x}  \varepsilon _{,x}   
-2  \frac{\partial ^{2}T}{\partial \rho \partial \gamma} \rho_{,x}  \gamma_{,x} \right.\allowdisplaybreaks\\
&\qquad\left.- \frac{\partial ^{2}T}{\partial \varepsilon ^{2}}  \varepsilon _{,x}^{2}  
-2 \frac{\partial ^{2}T}{\partial \varepsilon \partial \gamma}  \varepsilon _{,x}  \gamma_{,x}  
-\frac{\partial ^{2}T}{ \partial \gamma^{2}}  \gamma_{,x}^{2} \right)\allowdisplaybreaks\\
&\quad-\Lambda^{(1)}_3\left(
( v\rho_{,x}  +\rho   v_{,x})\varepsilon _{,x}   
-\left(\frac{\partial T}{\partial \rho }   \rho_{,x}  
+ \frac{\partial T}{\partial \varepsilon }  \varepsilon _{,x}    
+\frac{\partial T}{\partial \gamma}  \gamma_{,x} \right)   v_{,x}\right.\allowdisplaybreaks\\
&\qquad\left.+\frac{\partial ^{2}q}{\partial \rho ^{2}}  \rho_{,x}^{2}
+2  \frac{\partial ^{2}q}{\partial \rho\partial \varepsilon} \rho_{,x} \varepsilon _{,x}  
+2  \frac{\partial ^{2}q}{\partial \rho\partial\gamma}  \rho_{,x}\gamma_{,x} \right.\allowdisplaybreaks\\
&\qquad\left.+\frac{\partial ^{2}q}{\partial \varepsilon ^{2}}  \varepsilon _{,x}^{2}
+2\frac{\partial ^{2}q}{\partial \varepsilon \partial \gamma}  \varepsilon _{,x}  \gamma_{,x} 
+\frac{\partial ^{2}q}{\partial \gamma^{2}}  \gamma_{,x}^{2}  \right)\allowdisplaybreaks\\
&\quad-\Lambda^{(1)}_4\left((v \rho_{,x}+\rho   v_{,x})\gamma_{,x}    
+\frac{\partial ^{2}\phi }{\partial \rho ^{2}}   \rho_{,x}^{2}
+2  \frac{\partial ^{2}\phi }{\partial \rho \partial \varepsilon}   \rho_{,x} \varepsilon _{,x} \right.\allowdisplaybreaks\\
&\qquad\left.+2\frac{\partial ^{2}\phi }{\partial \rho \partial \gamma}  \rho_{,x} \gamma_{,x}    
+\frac{\partial ^{2}\phi }{ \partial \varepsilon ^{2}}  \varepsilon _{,x}^{2} 
+2  \frac{\partial ^{2}\phi }{ \partial \varepsilon \partial \gamma}  \varepsilon _{,x}  \gamma_{,x}  
+\frac{\partial ^{2}\phi }{\partial \gamma^{2}}  \gamma_{,x}^{2} \right)\ge 0,
\end{align*}
where we can distinguish the \emph{highest derivatives}, say
\[
\{\rho_{,t},v_{,t},\varepsilon_{,t},\gamma_{,t},\rho_{,tx},v_{,tx},\varepsilon_{,tx},\gamma_{,tx},\rho_{,xxx},v_{,xxx},\varepsilon_{,xxx},\gamma_{,xxx}\},
\]
and the \emph{higher derivatives}, say
\[
\{\rho_{,xx},v_{,xx},\varepsilon_{,xx},\gamma_{,xx}\}.
\]
As expected, the entropy inequality is linear in the highest derivatives and quadratic in the higher ones; 
the coefficients are at most functions of the field and the state variables.
By vanishing the coefficients of the highest derivatives, we determine the Lagrange multipliers, say
\begin{equation}
\begin{aligned}
&\Lambda_1^{(0)}=\rho\frac{\partial s}{\partial\rho}, \qquad 
&&\Lambda_2^{(0)}=-\frac{\rho_{,x}}{\rho}\frac{\partial s}{\partial v_x},\\
&\Lambda_3^{(0)}=\frac{\partial s}{\partial\varepsilon}-\frac{\rho_{,x}}{\rho}\frac{\partial s}{\partial\varepsilon_{,x}}, \qquad 
&&\Lambda_4^{(0)}=\frac{\partial s}{\partial\gamma}-\frac{\rho_{,x}}{\rho}\frac{\partial s}{\partial\gamma_{,x}},\\
&\Lambda_1^{(1)}=\rho\frac{\partial s}{\partial \rho_{,x}},\qquad
&&\Lambda_2^{(1)}=\frac{\partial s}{\partial v_{,x}},\\
&\Lambda_3^{(1)}=\frac{\partial s}{\partial \varepsilon_{,x}},\qquad
&&\Lambda_4^{(1)}=\frac{\partial s}{\partial \gamma_{,x}},
\end{aligned}
\end{equation}
as well as the following restrictions involving the entropy, the Cauchy stress tensor, the heat flux and the flux of internal variable:
\begin{equation}
\begin{aligned}
&\frac{\partial s}{\partial v_{,x}}\frac{\partial T}{\partial \rho_{,x}}-
\frac{\partial s}{\partial \varepsilon_{,x}}\frac{\partial q}{\partial \rho_{,x}}-
\frac{\partial s}{\partial \gamma_{,x}}\frac{\partial \phi}{\partial \rho_{,x}}=0,\\
&\frac{\partial s}{\partial v_{,x}}\frac{\partial T}{\partial v_{,x}}-
\frac{\partial s}{\partial \varepsilon_{,x}}\frac{\partial q}{\partial v_{,x}}-
\frac{\partial s}{\partial \gamma_{,x}}\frac{\partial \phi}{\partial v_{,x}}=0,\\
&\frac{\partial s}{\partial v_{,x}}\frac{\partial T}{\partial \varepsilon_{,x}}-
\frac{\partial s}{\partial \varepsilon_{,x}}\frac{\partial q}{\partial \varepsilon_{,x}}-
\frac{\partial s}{\partial \gamma_{,x}}\frac{\partial \phi}{\partial \varepsilon_{,x}}=0,\\
&\frac{\partial s}{\partial v_{,x}}\frac{\partial T}{\partial \gamma_{,x}}-
\frac{\partial s}{\partial \varepsilon_{,x}}\frac{\partial q}{\partial \gamma_{,x}}-
\frac{\partial s}{\partial \gamma_{,x}}\frac{\partial \phi}{\partial \gamma_{,x}}=0.
\end{aligned}
\end{equation}
The latter, joined with the conditions obtained by annihilating the coefficients of the linear terms  in the higher derivatives, provide the restrictions on the constitutive functions. These conditions can be solved so that we are able to provide an explicit solution that is proved to satisfy the remaining restrictions expressed as inequalities.
To proceed further, we start by taking the specific entropy as the sum of an equilibrium term and a non-equilibrium part expressed as a  
quadratic form in the gradients entering the state space; this quadratic part must be semidefinite negative in order to verify the principle of maximum entropy at equilibrium. By using some routines written in the Computer Algebra System Reduce \cite{Reduce}, we obtain the following solution to all the  thermodynamic restrictions:
\begin{equation}\label{entr_1}
\begin{aligned}
s&=s_0(\rho,\varepsilon)+s_1(\rho,\gamma)\rho_{,x}^2,\\
T&=\rho^2\frac{\partial s_0}{\partial \rho}\left(\frac{\partial s_0}{\partial\varepsilon}\right)^{-1}
+\tau_1(\rho,\varepsilon,\gamma)v_{,x}\\
&+\rho^2\left(\frac{\partial s_0}{\partial\varepsilon}\frac{\partial s_1}{\partial\gamma}\right)^{-1}
\left(s_1\frac{\partial^2 s_1}{\partial\rho\partial\gamma}-\frac{\partial s_1}{\partial\rho}\frac{\partial s_1}{\partial\gamma}\right)\rho_{,x}^2\\
&+\rho^2\left(\frac{\partial s_0}{\partial\varepsilon}\frac{\partial s_1}{\partial\gamma}\right)^{-1}
\left(s_1\frac{\partial^2 s_1}{\partial\gamma^2}-2\left(\frac{\partial s_1}{\partial\gamma}\right)^2\right)\rho_{,x}\gamma_{,x},\\
q&= q_1(\rho,\varepsilon,\gamma)\varepsilon_{,x}+q_2(\rho,\varepsilon,\gamma)\rho_{,x}+q_3(\rho,\varepsilon,\gamma)v_{,x},\\
\phi&=\left(2\frac{\partial s_1}{\partial\gamma}\rho_{,x}\right)^{-1}
\left(q_2\frac{\partial s_0}{\partial \varepsilon}+f(\rho,\varepsilon,\gamma)+2k\frac{\partial s_1}{\partial\gamma}\rho_{,x} -2\rho^2s_1v_{,x}\right),\\
J_s&=q\frac{\partial s_0}{\partial\varepsilon}-\frac{1}{2}\left(q_2\frac{\partial s_0}{\partial \varepsilon}+f(\rho,\varepsilon,\gamma)-2\rho^2s_1 v_{,x}\right)\rho_{,x},
\end{aligned}
\end{equation}
where the function $s_1(\rho,\gamma)$ must be a negative function in order the principle of maximum entropy at the equilibrium be satisfied.
Moreover, the reduced entropy inequality becomes 
\begin{equation}
\label{reduced1}
\begin{aligned}
&(q_1\varepsilon_{,x}+q_2\rho_{,x}+q_3v_{,x})\left(\frac{\partial^2 s_0}{\partial \rho\partial\varepsilon}\rho_{,x}+\frac{\partial^2 s_0}{\partial\varepsilon^2}\varepsilon_{,x}\right)+\tau_1\frac{\partial s_0}{\partial\varepsilon} v_{,x}^2\\
&-\left(\frac{\partial s_1}{\partial\gamma}\right)^{\frac{1}{2}}\left(\frac{\partial g(\rho,\varepsilon)}{\partial \rho}\rho_{,x}+\frac{\partial g(\rho,\varepsilon)}{\partial \varepsilon}\varepsilon_{,x}\right)\rho_{,x}\geq 0,
\end{aligned}
\end{equation}
along with the constraint
\begin{equation}
q_2\frac{\partial s_0}{\partial\varepsilon}+f(\rho,\varepsilon,\gamma)-\left(\frac{\partial s_1}{\partial\gamma}\right)^{\frac{1}{2}}g(\rho,\varepsilon)=0,
\end{equation}
where $f(\rho,\varepsilon,\gamma)$ and $g(\rho,\varepsilon)$ are arbitrary functions of their arguments.

The inequality \eqref{reduced1} is satisfied if and only if the following conditions hold true:
\begin{equation}
\label{diffconstrgen}
\begin{aligned}
&q_1\frac{\partial^2 s_0}{\partial\varepsilon^2}\geq 0,\qquad\tau_1\frac{\partial s_0}{\partial\varepsilon} \geq 0,\\
&q_2\frac{\partial^2 s_0}{\partial \rho\partial\varepsilon}-\left(\frac{\partial s_1}{\partial\gamma}\right)^{\frac{1}{2}}\frac{\partial g}{\partial \rho}\geq 0,\qquad q_3^2\frac{\partial^2 s_0}{\partial\varepsilon^2}-4\frac{\partial s_0}{\partial\varepsilon}\tau_1 q_1\geq 0,\\
&\left(q_3^2\frac{\partial^2 s_0}{\partial\rho \partial\varepsilon}-4\frac{\partial s_0}{\partial\varepsilon}\tau_1 q_2\right)\frac{\partial^2 s_0}{\partial\rho \partial\varepsilon}+4\left(\frac{\partial s_1}{\partial\gamma}\right)^{\frac{1}{2}}\frac{\partial s_0}{\partial\varepsilon}\frac{\partial g}{\partial \rho}\tau_1\leq 0,\\
&\left(q_2\frac{\partial^2 s_0}{\partial\varepsilon^2}-q_1\frac{\partial^2 s_0}{\partial\rho \partial\varepsilon}\right)^2+4\left(\frac{\partial s_1}{\partial\gamma}\right)^{\frac{1}{2}}\frac{\partial^2 s_0}{\partial\varepsilon^2}\frac{\partial g}{\partial\rho} q_1+\frac{\partial s_1}{\partial\gamma}\left(\frac{\partial g}{\partial\varepsilon}\right)^2\\
&-2\left(\frac{\partial s_1}{\partial\gamma}\right)^{\frac{1}{2}}\left(q_2\frac{\partial^2 s_0}{\partial\varepsilon^2}+q_1\frac{\partial^2 s_0}{\partial\rho \partial\varepsilon}\right)
\frac{\partial g}{\partial\varepsilon}\leq 0,\\
&\left(q_2\frac{\partial^2 s_0}{\partial\varepsilon^2}-q_1\frac{\partial^2 s_0}{\partial\rho \partial\varepsilon}\right)^2\frac{\partial s_0}{\partial\varepsilon}\tau_1-\left(\frac{\partial s_1}{\partial\gamma}\right)^{\frac{1}{2}}\left(q_3^2\frac{\partial^2 s_0}{\partial\varepsilon^2}-4\frac{\partial s_0}{\partial\varepsilon}\tau_1 q_1\right)\frac{\partial^2 s_0}{\partial\varepsilon^2}\frac{\partial g}{\partial\rho}\\
&-\left(\frac{\partial s_1}{\partial\gamma}\right)^{\frac{1}{2}}\left(q_3^2\frac{\partial^2 s_0}{\partial \rho\partial\varepsilon}\frac{\partial^2 s_0}{\partial\varepsilon^2}-2\left(q_2\frac{\partial^2 s_0}{\partial\varepsilon^2}+q_1\frac{\partial^2 s_0}{\partial \rho\partial\varepsilon}\right)\frac{\partial s_0}{\partial \varepsilon}\tau_1\right)\frac{\partial g}{\partial \varepsilon}\\
&+\frac{\partial s_1}{\partial \gamma}\frac{\partial s_0}{\partial\varepsilon}\left(\frac{\partial g}{\partial\varepsilon}\right)^2 \tau_1\leq 0.
\end{aligned}
\end{equation}

The solution so recovered contains some degrees of freedom that can be fixed in order to model specific physical situations. The constitutive equation \eqref{entr_1}$_1$ can be interpreted as an extension of the equilibrium constitutive equation to non-equilibrium  situations \cite{Muschik-Ehrentraut}.
In what follows, we limit ourselves to discuss in detail the case where $g(\rho,\varepsilon)=0$, whereupon the constraints 
\eqref{diffconstrgen} simplify as
\begin{equation}
\label{lastrestr1}
\begin{aligned}
&q_2\frac{\partial^2 s_0}{\partial \rho \partial\varepsilon}\geq 0,\qquad q_1\frac{\partial^2 s_0}{\partial\varepsilon^2}\geq 0,\qquad \tau_1\frac{\partial s_0}{\partial\varepsilon}\geq 0,\\
&\frac{\partial^2 s_0}{\partial \rho\partial\varepsilon}\left(4\frac{\partial s_0}{\partial\varepsilon}\tau_1 q_2-q_3^2\frac{\partial^2 s_0}{\partial \rho\partial\varepsilon}\right)\geq 0,\\
&\frac{\partial^2 s_0}{\partial\varepsilon^2}\left(4\frac{\partial s_0}{\partial\varepsilon}\tau_1 q_1-q_3^2\frac{\partial^2 s_0}{\partial\varepsilon^2}\right)\geq 0,
\end{aligned}
\end{equation}
together with 
\begin{equation}
\label{relationq1q2}
q_2\frac{\partial^2 s_0}{\partial\varepsilon^2}-q_1\frac{\partial^2 s_0}{\partial \rho\partial\varepsilon}=0.
\end{equation}

Some comments about the constitutive relations so characterized are in order. 
In equilibrium situations, in which the gradients of the field variables (except at most the density $\rho$) vanish, let us define the absolute temperature $\theta$ by the classical thermodynamical relation $\displaystyle \frac{1}{\theta}=\frac{\partial s_0}{\partial\varepsilon}$. 
Under the hypothesis of invertibility of 
$\theta$ with respect to $\varepsilon$, which is allowed by the positivity of the specific heat 
$\displaystyle c = \frac{\partial \varepsilon}{\partial\theta}$, the internal energy $\varepsilon$ can be
expressed as function of the arguments $\rho$ and $\theta$. Thus,  differentiating with respect to $\rho$ the
condition
\[
\frac{\partial s_0(\rho,\varepsilon(\rho,\theta))}{\partial \varepsilon}-\frac{1}{\theta}=0,
\]
we get
\begin{equation}
\label{sign}
\frac{\partial^2 s_0}{\partial\rho\partial\varepsilon}+\frac{\partial^2 s_0}{\partial \varepsilon^2}\frac{\partial\varepsilon}{\partial\rho}=0,
\end{equation}
that, used in \eqref{relationq1q2}, provides
\begin{equation}
q_2=-q_1\frac{\partial\varepsilon}{\partial\rho}.
\end{equation}

Thus, the heat flux reduces to
\begin{equation}
q = q_1\frac{\partial\varepsilon}{\partial\theta}\theta_{,x}+q_3v_x,
\end{equation}
\emph{i.e.}, when $q_3=0$, we have the classical Fourier law of heat conduction. 

Since 
\[
\frac{\partial s_0}{\partial\varepsilon}=\frac{1}{\theta}>0, \qquad\frac{\partial^2 s_0}{\partial\varepsilon^2}=-\frac{1}{\theta^2}\frac{\partial\theta}{\partial\varepsilon}< 0,
\]
it is $\tau_1\geq 0$, and, as physics prescribes,  $q_1\leq 0$.
As far as $q_2$ is concerned, its sign is the same as that of $\displaystyle\frac{\partial\varepsilon}{\partial\rho}$, which in turn, from \eqref{sign}, has the same sign as  
$\displaystyle \frac{\partial^2 s_0}{\partial\rho\partial\varepsilon}$; thus, due to $\displaystyle\frac{\partial\varepsilon}{\partial\rho}\geq 0$, it is $q_2\geq 0$. 
Finally, the last two inequalities in \eqref{lastrestr1} provide
\begin{equation}
\label{onlyone}
q_3^2\leq 4 \left(\frac{\partial^2 s_0}{\partial \varepsilon^2}\right)^{-1}\frac{\partial s_0}{\partial\varepsilon}q_1\tau_1.
\end{equation}
Last, but not the least, it is worth of being observed that the entropy flux contains the classical term $\displaystyle \frac{q}{\theta}$  and an additional term (extra-flux) coming from the application of the procedure  without the need of postulating it at the beginning. Finally, at equilibrium the flux $\phi$ of the internal variable $\gamma$ reduces to a constant, and
the stress tensor $T$ assumes the classical local form if the mass density is constant. 

\subsection{Korteweg fluids}
Here, we consider the case of a fluid of grade 3 \cite{Truesdell_Noll}; in fact,  we include in the state space the second spatial derivative of the mass density \cite{Dunn-Serrin}. For this class of fluids, Korteweg 
\cite{Korteweg} proposed the Cauchy stress tensor to be given by a constitutive equation like
\begin{equation} 
\label{kort}
T_{ij}=\left(-p+\sum_{k=1}^3\left(\alpha_1\frac{\partial^2\rho}{\partial x_k^2}+\alpha_2\frac{\partial\rho}{\partial x_k}
\frac{\partial\rho}{\partial x_k}\right)\right)\delta_{ij}+\alpha_3\frac{\partial\rho}{\partial x_i}
\frac{\partial\rho}{\partial x_j}+\alpha_4\frac{\partial^2\rho}{\partial x_i\partial x_j},
\end{equation}
where $\rho$ denotes the mass density, $p$  the pressure of the fluid, and $\alpha_i$, $i=1,\ldots,4$,  
suitable material functions depending on density and temperature. These fluids received a moderate attention in the 
literature after the pioneering paper by Dunn and Serrin \cite{Dunn-Serrin}, where the compatibility with the basic tenets 
of rational continuum thermodynamics \cite{Truesdell} has been extensively studied. They have been studied also in 
\cite{CST-JMP-2009,COT-JMP-2011,CST-JNET-2010,COP-Elasticity-2011} 
by means of an extended Liu procedure, and by 
Heida and M\'alek \cite{Heida-Malek} following a different methodology. 

By limiting to the one-dimensional case, the balance equations read 
\begin{equation}
\label{korteweg}
\begin{aligned}
&\mathcal{E}_1\equiv\frac{D\rho}{D t} + \frac{D (\rho v)}{D x}=0, \\
&\mathcal{E}_2\equiv\frac{D(\rho v)}{D t} + \frac{D (\rho v^2  - T)}{D x}=0,\\
&\mathcal{E}_3\equiv\frac{D}{D t}\left(\rho\varepsilon+\rho \frac{v^2}{2}\right) +\frac{D }{D x}\left(\rho v\varepsilon+\rho \frac{v^3}{2}-Tv+ q\right)=0,
\end{aligned}
\end{equation}
where $\rho$ is the mass density, $v$ the velocity, and $\varepsilon$ the internal energy per unit mass;
moreover, the stress $T$ and the heat flux $q$ must be assigned by means of constitutive equations.
The constitutive equations must be such that for every admissible process the entropy inequality 
\begin{equation}
\rho\left(\frac{Ds}{D t}+v\frac{Ds}{Dx}\right)+ \frac{DJ_s}{Dx} \ge 0,
\end{equation}
being $s$ the specific entropy and $J_s$ the entropy flux, is satisfied.

Let us assume the state space spanned by
\begin{equation}
\mathcal{Z}\equiv\{\rho, \varepsilon, \rho_{,x}, \varepsilon_{,x}, v_{,x}, \rho_{,xx}\}.
\end{equation}

In this case, the exploitation of second law of thermodynamics requires that we take into account the constraints 
imposed on the thermodynamic processes by the balance equations together with their first and second order spatial derivatives; 
nevertheless, we observe that, since the unique second order spatial derivative belonging to the state space is $\rho_{,xx}$, the unique second order extension we need to use as constraint is $\displaystyle\frac{D^2\mathcal{E}_1}{Dx^2}$.
Thus, the entropy inequality writes:
\begin{equation}
\label{entropyconstrainedkorteweg}
\begin{aligned}
&\rho\left(\frac{Ds}{D t}+v\frac{Ds}{Dx}\right)+ \frac{DJ_s}{Dx} \\
&\quad- \Lambda^{(0)}_1 \mathcal{E}_1- \Lambda^{(0)}_2 \mathcal{E}_2- \Lambda^{(0)}_3 \mathcal{E}_3\\
&\quad-\Lambda^{(1)}_1\frac{D\mathcal{E}_1}{Dx}-\Lambda^{(1)}_2\frac{D\mathcal{E}_2}{Dx}
-\Lambda^{(1)}_3\frac{D\mathcal{E}_3}{Dx}-\Lambda^{(2)}_1\frac{D^2\mathcal{E}_1}{Dx^2} \geq 0.
\end{aligned}
\end{equation}

By expanding the derivatives in \eqref{entropyconstrainedkorteweg}, we obtain a huge expression that we omit to report here;  
it results linear in the  highest derivatives, say
\[
\{\rho_{,t},v_{,t},\varepsilon_{,t},\rho_{,tx},v_{,tx},\varepsilon_{,tx},\rho_{,txx},v_{,txx},\varepsilon_{,txx},v_{,xxxx},\varepsilon_{,xxxx},\rho_{,xxxxx}\},
\]
and cubic in  the higher derivatives, say
\[
\{v_{,xx},\varepsilon_{,xx},\rho_{,xxx},v_{,xxx},\varepsilon_{,xxx},\rho_{,xxxx}\}.
\]

To proceed further, let us write the specific entropy as the sum of the equilibrium part  defined for homogeneous states and
a semidefinite negative quadratic form (in order to satisfy the principle of maximum entropy at the equilibrium) in the gradients appearing in the state space.  At this stage we do not specify the form of the entropy flux.

The restrictions imposed by entropy inequality can be solved and provide the following solution:
\begin{equation}
\begin{aligned}\label{entr_gen}
&s = s_0(\rho,\varepsilon)+s_1(\rho)\rho_{,x}^2,\\
&q = q_1(\rho,\varepsilon)\varepsilon_{,x}+q_2(\rho,\varepsilon)\rho_{,x}+q_3(\rho,\varepsilon)v_{,x},\\
&T = \rho^2\left(\frac{\partial s_0}{\partial \varepsilon}\right)^{-1}\left(\frac{\partial s_0}{\partial \rho}-\frac{\partial s_1}{\partial \rho}\rho_{,x}^2-2 s_1\rho_{,xx}\right)+\tau_1(\rho,\varepsilon)v_{,x},
\end{aligned}
\end{equation}
where
\begin{equation}
\label{constrkdv}
\begin{aligned}
&q_1\le 0, \qquad q_2\ge 0, \qquad \tau_1\ge 0,\\
&q_2\frac{\partial^2 s_0}{\partial\varepsilon^2}-q_1\frac{\partial^2 s_0}{\partial\rho\partial\varepsilon}=0,\\
&4\frac{\partial s_0}{\partial\varepsilon}\tau_1 q_2-q_3^2\frac{\partial^2 s_0}{\partial\rho\partial\varepsilon}\geq 0,\\
&4\frac{\partial s_0}{\partial\varepsilon}\tau_1 q_1-q_3^2\frac{\partial^2 s_0}{\partial\varepsilon^2}\leq 0,
\end{aligned} 
\end{equation}
and $s_1\leq 0$ in order the principle of maximal entropy at the equilibrium be satisfied.
Finally, the entropy flux turns out to be
\begin{equation}
\begin{aligned}\label{flux_entr}
J_s&=\frac{\partial s_0}{\partial\varepsilon}\left(q_1\varepsilon_{,x}+q_2\rho_{,x}+q_3v_{,x}\right)+2\rho^2 s_1\rho_{,x}v_{,x}=\\
&=\frac{q}{\theta}+2\rho^2 s_1\rho_{,x}v_{,x},
\end{aligned}
\end{equation}
where $\theta$, defined as usual as
\begin{equation}
\frac{1}{\theta}=\frac{\partial s_0}{\partial\varepsilon},
\end{equation}
is the absolute temperature.

Let us observe that, in the expression \eqref{flux_entr}  of entropy flux, the second contribution  represents  the  entropy extra-flux \cite{Muller}, related to the gradients of density and velocity.
Moreover, a similar reasoning as in previous subsection shows that it is $\displaystyle q_2=-q_1\frac{\partial\varepsilon}{\partial\rho}$ so that, choosing $q_3=0$, we recover the classical Fourier law for heat flux. Also, the last two inequalities 
in \eqref{constrkdv} provide relation \eqref{onlyone}.
Finally, if we consider the classical case, \emph{i.e.}, $s = s_0(\rho,\varepsilon)$, the stress tensor $T$ is expressed in local form, and the classical constitutive equation for the entropy flux $\displaystyle J_s=\frac{q}{\theta}$ is recovered.

Therefore,  in order to obtain a constitutive equation for the stress tensor containing first and second order derivatives of mass density we need to assume that $s$ depends on the gradient of $\rho$, and the entropy flux involves an extra-flux.
As a last remark, a stress tensor depending on the gradients of the density can be obtained only if we use the extended
procedure for the exploitation of entropy inequality \cite{CGOP-miscele-2020}.

\section{Conclusions}
\label{sec:conclusions}
In this paper, we discussed the extended Liu procedure in order to investigate the restrictions placed by an entropy inequality
on the constitutive equations of a continuum whose state space contains spatial derivatives of the unknown fields. The analysis is
performed first from a purely mathematical viewpoint by considering a system of balance laws sufficiently general to contain
the governing equations of continua, and sufficient conditions for the fulfilment of an entropy-like inequality have been derived. Then, we specialized the results to two physical cases with first or second order  non-local 
constitutive equations. Remarkably, in the application of the procedure we do not modify the energy balance equation with the  inclusion of extra-terms (like the interstitial  working), and we do not prescribe \emph{a priori} the form of entropy flux, whose expression arises as a result of
the method; in the applications we considered, the procedure provides an entropy flux decomposed in the classical term and an extra-flux.

We limited ourselves to one-dimensional models, and in the considered physical applications we were able to solve the
constraints imposed by the exploitation of the entropy inequality, so determining an explicit expression of the constitutive functions
by assuming  an expansion of the specific entropy at first order in the squared gradients of the field variables entering the state space.

The procedure in principle allows us to fix the constitutive equations (for stress tensor, heat flux, \ldots), according to experiments, and determine the form of the entropy flux algorithmically. 
It is trivial to observe that the procedure requires a huge amount of computation, increasing with the order of non-localities;
however, such computations can be almost automatically carried out by using a Computer Algebra System.

Work is in progress about the investigation of the extended Liu procedure for multi-dimensional continuous media, where
other general principles of representation theory \cite{Smith3} of vectorial and tensorial quantities need to be considered. 

\section*{Acknowledgments}
The authors acknowledge partial supported by G.N.F.M. of ``Istituto Nazionale di Alta Matematica'' and University of Messina. The authors gratefully thank dr. Luca Amata for drawing their attention to the paper \cite{Mishkov}.

\end{document}